\documentclass[useAMS,usenatbib,a4]{mn2e}  

\usepackage[dvips]{graphicx}
\usepackage{natbib}

\title[Spectral synthesis including massive binaries ]{Spectral
  population synthesis including massive binaries}
\author[Eldridge \& Stanway]{John J. Eldridge$^{1}$\thanks{jje@ast.cam.ac.uk} \& Elizabeth R. Stanway$^2$\\
  $^1$ Institute of Astronomy, University of Cambridge, Madingley Road, Cambridge, CB3 0HA, UK.\\
  $^2$ H H Wills Physics Laboratory, Tyndall Avenue, Bristol, BS8 1TL, UK\\
}

\pagerange{\pageref{firstpage}--\pageref{lastpage}} \pubyear{2009}
\begin{document}
\maketitle
\label{firstpage}

\begin{abstract}
  We have constructed a new code to produce synthetic spectra of
  stellar populations that includes massive binaries. We have tested
  this code against the broadband colours of unresolved young massive
  stellar clusters in nearby galaxies, the equivalent widths of the
  Red and Blue Wolf-Rayet bumps in star-forming SDSS galaxies and the
  UV and optical spectra of the star forming regions Tol-A and B in
  NGC5398. In each case we find a good agreement between our models
  and observations. We find that in general binary populations are
  bluer and have fewer red supergiants, and thus significantly less
  flux in the I-band and at longer wavelengths, than single star
  populations. Also we find that Wolf-Rayet stars occur over a wider
  range of ages up to $10^7$ years in a stellar population including
  binaries, increasing the UV flux and Wolf-Rayet spectral features at
  later times. In addition we find that nebula emission contributes
  significantly to these observed properties and must be considered
  when comparing stellar models with observations of unresolved
  stellar populations. We conclude that incorporation of massive
  stellar binaries can improve the agreement between observations and
  synthetic spectral synthesis codes, particularly for systems with
  young stellar populations.
\end{abstract}

\section{Introduction}

While bright individual stars can be resolved by observations of local
galaxies, for more compact or distant systems only the ensemble
properties of an unresolved population can be measured.  Through
necessity, the analysis applied in these cases differs. For resolved
populations it is possible to study each star in detail, and to
compare number ratios of different stellar types. By contrast, in
unresolved populations the relative contributions of different stellar
types are estimated from the overall spectral energy distribution and
from specific emission/absorption lines by using synthetic colours and
spectra calculated from spectral population synthesis (SPS)
codes. However SPS codes suffer from many uncertainties due to limited
modelling of contributions from numerically small, but nonetheless
important stellar sub-types, and due to different assumptions and
techniques for balancing the contribution of different components in
the population \citep{uncertain}. Binary evolution, and in particular
the effects of binary evolution on massive, short-lived but
ultraviolet luminous stars such as the Wolf-Rayet population, has been
neglected in existing SPS codes. 

An understanding of the signatures of massive stars in unresolved
young populations is essential for correctly interpreting the star
formation history and environment of starburst regions both locally
\citep[see][]{schavacca,brinchmann2} and at high redshifts where
starbursts are more prevalent, and data - both photometric and
spectroscopic - sparser. At the highest redshifts, the predicted
signatures of massive Population III stars
\citep[e.g.][]{2002A&A...382...28S} must be disentangled from those of
massive stars at higher metallicities, and so the latter must first be
well understood. In this paper we address the inclusion of binary
evolution paths in spectral synthesis codes, and explore their effects
on fitting of unresolved stellar spectra of star-forming galaxies.

\citet{EIT} demonstrated that the predicted properties of stellar
populations including binary models matched those of resolved stellar
populations better than single-star models. Independently
\citet{brinchmann2} showed that the same binary models provide a good
fit to the stellar populations of unresolved Wolf-Rayet galaxies in
Sloan Digital Sky Survey (SDSS) data. Their analysis derived the
relative numbers of O stars and Wolf-Rayet stars from the equivalent
widths of certain spectral features in the optical regime, and
compared this ratio with those modelled. A more direct comparison is
possible: between the observed spectra and synthetic spectra derived
from a model binary population.

Codes to create such synthetic spectra exist, the most widely-used
being \textit{starburst99} \citep{1999ApJS..123....3L}. However few
such codes take into consideration the full effect of binary
evolution. \citet{vanbev} and \citet{belkus} and references therein
included binary evolution in a rapid population synthesis code. Their
binary model was based on 1000 detailed stellar
models, while \citet{otherbinsynth} adapted the same binary models to
be included in \textit{starburst99}. However since the single star and
binary models came from different stellar evolution codes they
modified the binary models to match with the single star evolutionary
tracks in \textit{starburst99}. We also note that \citet{hanbin}
incorporated binary evolution of low-mass stars to explain the
observed UV flux in some elliptical galaxies. However they have not
extended this to higher mass binaries.

Our self-consistent approach uses full sets of detailed single stars
and binary models from \citet{EIT}. Both were created with one code,
the Cambridge STARS code, with identical input physics. Also the
evolution of our binary population is based on 3000 detailed models
for each metallicity.

In addition our spectral synthesis code presented here incorporates
theoretical rather than empirical stellar spectral libraries wherever
possible in order to provide as complete a theoretical model as
possible to compare to observational data. Therefore the only empirical
results that go into our code are the equivalent widths of HeII lines
for Of stars and a subset of WNL stars, the strength of convective overshooting,
the mixing length for convection and the mass-loss rates for red
supergiants and Wolf-Rayet stars.

Finally, given this theoretical bias, we account for nebular emission
from the gas and dust surrounding our stellar population through use of
the photoionization program \textsc{Cloudy} \citep{cloudy}. In
combination with the purely synthetic stellar spectra, this leads to
an accurate model of the total emission from stars, gas and dust in a
model galaxy or star cluster.

The structure of this paper is as follows: In section
\ref{sec:synthetic_spectra} we detail the construction of our
synthetic stellar spectra, and how we process this output with
\textsc{Cloudy}. In section \ref{sec:verif-local-appl} we compare the
broad-band magnitudes and colours predicted from our models to
observed young massive clusters in nearby galaxies (section
\ref{sec:WR_in_M31}), to equivalent line widths for Wolf-Rayet
features in unresolved SDSS galaxies (section \ref{sec:sdss_gals}),
and to the observed UV and optical spectra of the Tol A \& B regions
discussed by \citet[][ section \ref{tolab}]{sidoli}. Finally we
discuss the implications and interpretation of our results and list
our conclusions.

\section{Synthetic spectra of stellar populations}
\label{sec:synthetic_spectra}

Creating a synthetic spectra of a stellar population requires the
combination of stellar evolution models, model stellar atmospheres and
a nebular emission model. In this section we detail the source of
these sets of models and how they are combined to produce a composite
population spectrum.

\subsection{Stellar evolution models}
\label{sec:stellar_models}

We use stellar models from the Cambridge STARS code \citep[][ and
  references therein]{egg,EIT}, specifically those calculated in
\citet{EIT}. Their key feature is that there is not only a set of
detailed single star models, but also an extensive set of detailed
binary star models which are key to producing a realistic synthetic
stellar population. We consider stellar models at five different
metallicities: Z=0.001, 0.004, 0.008, 0.020 and 0.040 (where a
metallicity of Z=0.020 is conventionally considered solar), with
hydrogen mass fraction, $X=0.75-2.5Z$, and helium mass fraction,
$Y=0.25+1.5Z$. We use the method described in \citet{EIT} to model the
primary and secondary stars in a binary and to account for
mergers. The only difference in our current procedure is that at each
timestep we now select a model atmosphere that best represents the
model and combine these to form a composite spectrum for the
population. Given that stellar evolution is non-linear and binary
evolution is even less predictable, we do not interpolate between
models with different masses and initial binary parameters, but rather
weight each stellar model by a Salpeter initial mass function
(IMF). We note that the results presented in \citet{EIT} are for
continuous star formation. Here we consider the evolution of a single
instantaneous burst with age, and do not consider multiple bursts of
star formation. Features in the rest-UV are sensitive primarily to
young stars so are relatively unaffected by older underlying stellar
populations, while such an older stellar population will tend to boost
the optical continuum, and hence reduce the strength of line emission
features in this region.

\subsubsection{Description of the binary models and population synthesis}
\label{sec:binary-mod}

While a full description of our details models can be found in
\citet{EIT} we provide a brief overview. We have modified our stellar
evolution code to model binary evolution.  The details of our binary
interaction algorithm are relatively simple compared to the scheme
outlined in, for example, \citet{HPT02}. We make a number of
assumptions in producing our code to keep it relatively simple. Our
aim was to investigate the effect of enhanced mass loss due to binary
interactions on stellar lifetimes and populations; therefore we
concentrated on these aspects rather than incorporating additional
physical processes, each of which would add more free parameters to
our models and potentially associated uncertainties on those
parameters or the mechanisms concerned. For example we do not include
wind accretion, gravitational breaking or magnetic breaking. The
processes are unlikely to be important in the evolution of massive
binaries due to the short evolutionary timescales of massive
stars. Processes that would have a more significant results on our
binary population are that we employ a simple tidal model that assumes
stellar rotation only becomes synchronous with the orbit during
mass-transfer events and that all the orbits are circular throughout
their evolution. \citet{HPT02} and \citet{2009MNRAS.396.1699S} find
that including accurate treatments of these do not provide new
evolutionary paths, but do alter the initial separation at which
different evolutionary paths occur. For example inclusion of an
accurate tide model increases the maximum initial separation for a
mass-transfer event to occur in a binary. This suggests we could be
underestimating the number of binary interactions in our stellar
models.

We also make assumptions in calculating our synthetic population to
avoid calculating a large number of models.  For example, we do not
model accretion onto the secondary in the detailed code. We take the
greater of the final mass of the secondary at the end of the primary
code or its initial mass when we consider the evolution of the
secondary. This avoids calculating 10 times more secondary models than
primary models but possibly missed important effects of accretion on
the evolution.

We always treat the primary as the initially more massive star and we
only evolve one star at a time with our detailed code. Our models have
initial separations that take values between $\log(a/R_{\odot})=1$ to
4 in steps of 0.25 dex. The mass ratio takes values of $q=0.1$, 0.3,
0.5, 0.7 and 0.9. The primary initial masses range from 5 to
120\,M$_{\odot}$, so the smallest mass star in our population is
0.5\,M$_{\odot}$. However we only include a star's contribution to our
synthetic spectrum if its mass is above 5\,M$_{\odot}$ at any point of
its evolution. Because of this constraint, we do not attempt to model
populations older than approximately 40\,Myrs old.

When we evolve the primary in detail, it has a shorter evolutionary
time-scale than the secondary which remains on the main sequence until
after the primary completes its evolution and so we can determine the
state of the secondary using the single stellar evolution equations of
\citet{HPT00}. When we evolve the secondary in detail, we assume that
its companion is the compact remnant of the primary (a white dwarf,
neutron star or black hole) and treat this as a point mass. If we find
that the binary systems experience a merger we use a very simple merger
scheme in which, when the stars come into contact, all the mass of the
secondary is accreted onto the primary. Then subsequent evolution
occurs as a single star.

Our population synthesis calculations are built upon those described
in \citet{EIT} with some improvements. The important point of
evolution is what happens after the first supernova (SN) in a
binary. If a star has a carbon/oxygen core mass greater than $1.38 \,
{\rm M}_{\odot}$ and the final mass of the star is greater than
2\,M$_{\odot}$ we assume it explodes in a SN. If a neutron star is
formed, we determine the fate of the binary by using the work of
\citet{nskick} with the latest determination for the kick velocity
distribution from observations by \citet{newkick}. If the remnant is a
black hole, we assume that it receives a similar kick. Because the
masses of black holes are greater than neutron stars we use the kick
distribution of \citet{newkick} to fix the momentum distribution. We
calculate the resulting black hole kick velocity from $v_{\rm BH}
=v_{\rm NS} (1.4/M_{\rm BH})$.

For each primary model there are many possible outcomes after the
first SN. The binary might be unbound or remain bound. In the latter
case there are a range of different orbital separations possible,
dependant on the strength and direction of the SN kick. To determine
how important the different possible outcomes are we generate a random
SN kick velocity and direction, calculate the effect on the binary,
then repeat many times to estimate the relative importance of the
outcomes. This process leaves us with the weights to apply to our
secondary models when we include them in our synthetic population and
spectra.

\subsubsection{The effect of rotation}

Modelling binary systems provides in general similar effects to
including rotation in single star models. Both are very complex to
implement and currently there are few codes that include both. To our
knowledge only the code described by \citet{cantiello} and \citet{demink} (and references
there-in) does so, and is extremely computationally and human input
intensive. Trying to separate out the effects of binaries and
single-star rotation on stellar populations is difficult as both have
similar effects, that is to produce more Wolf-Rayet stars at the
expense of red-supergiants \citep{vaz07,EIT}

Both can also increase the number of massive main-sequence stars
observed. Rotation does this by increasing the amount of mixing during
the main sequence and thus extending lifetimes and making stars of the
same mass and luminosity have higher surface temperatures
\citep{mm2005}. For a detailed comparison of our single-star models to
the Geneva rotation models see \citet{evink}. Binaries can also
increase the number of main-sequence stars, due to secondaries
accreting material during binary interactions and becoming more
massive. However there is evidence that in such processes rotation may
be important \citep[e.g.][]{cantiello,2009MNRAS.396.1699S}. The
difference between the two processes however is that the enhancement
by binaries can be delayed to much later times than rotation. In
addition for rotation to have a significant effect all stars would
need to have initial rotation velocities around $300{\rm km \,
  s^{-1}}$.

We omit rotation from our models, probing instead the degree to which
binarity alone can explain the observed features of stellar
populations. Discrepancies between observation and our theoretical
models may indicate the role played by rotation.

\subsection{Model atmospheres}
\label{sec:model_atmospheres}

At each timestep, the properties of the stellar evolution model are
used to select the most appropriate theoretical stellar atmosphere
model from one of three sources. 

Firstly, for stars with hydrogen envelopes and surface temperatures
$<$25kK, we use the widely employed BaSeL V3.1 model atmosphere library
\citep{basel}. Stars with higher surface temperatures are treated as
OB stars. For these we use the high-resolution versions of the models
of \citet{crow}\footnote{Which can be found at \\
\texttt{http://zuserver2.star.ucl.ac.uk/$\sim$ljs/starburst/BM\_models/}}. Both
libraries are arranged in a grid over effective temperature and
surface gravity. We interpolate within this grid linearly to obtain
appropriate spectra for our models.

The most important difference between this work and previous analyses
is our use of the theoretical atmospheres of the Potsdam group
\citep{wn} for Wolf-Rayet stars (defined as having surface hydrogen
mass fraction $X \le 0.4$ and $\log(T_{\rm eff}/{\rm K}) \ge
4$). These are advanced atmosphere models that can be related directly
to stellar evolution models. We use not only the publicly available
models for WN stars but also a set of WC models from the same group
that are preliminary results (W.-R. Hamann \& A. Barniske, private
communication). We have compared the atmosphere models to low
resolution spectra produced by \citet{crow} and find broadly similar
results to those from these more detailed models. These represent an
important step forwards to producing a \textit{solely theoretical}
synthetic population spectrum rather than one based on difficult to
interpret empirical observations. These models are on a grid of
transformed radius and effective temperature which we interpolate
linearly between the values of our stellar models.

We use the Potsdam WR atmosphere models in the parameter space they
cover when $X \le 0.2$ and $\log(T_{\rm eff}/{\rm K}) \ge 4.45$). We
use the WNL models when $0.2 \ge X \ge 0.1$. When $0.1 \ge X \ge 0.01$
we interpolate between the WNL models and the WNE models. We use the
WNE models alone when $X \le 0.01$.

To determine when to switch to WC models we use the variable
$\alpha=(x_{\rm C}+x_{\rm O})/y$ where $x_{\rm C}$, $x_{\rm O}$ and
$y$ are the abundance by number of carbon, oxygen and helium
respectively. When $\alpha > 0.01$ we begin to interpolate between the
WNE and WC atmosphere models until $\alpha>0.26$. This value is chosen
as it is the value of the composition of the atmosphere model. Above
this value we use the WC atmosphere model alone.

This scheme omits a subset of stars that should be included as WR
stars, those with $4 \le \log(T_{\rm eff}/{\rm K}) \le 4.45$ and $0.2
\le X \le 0.4$. For these stars no model WR star spectra exist, so we
use the corresponding BaSeL or OB spectra but modified to include the
line luminosities of the HeII line at 1640\AA\ and the WR blue bump
contributed by these stars. We use the empirical line luminosities
given in \citet[][based on Crowther \& Hadfield, 2006]{brinchmann1}
for WNL stars rather than the older values in \citet{schavacca}. We
only apply this correction if the star has a luminosity above
$\log(L/L_{\odot})\ge 4.9$ to prevent a large contribution to these
features due to lower mass stars at late times. By using fixed line
luminosities given in Table \ref{brinchman_lums} at our assumed
luminosity limit, 10 percent of the total stellar emission is in these
lines. This proportion would be larger if we included these line
luminosities for less luminous stars. In \citet{EIT} it was found that
this luminosity limit was necessary to reproduce the observed ratio of
the number of WC to WN stars. In Table \ref{lineluminosities} we show
the mean line luminosities predicted for all WR stars in our synthetic
population. We find that in general we predict slightly lower mean
line luminosities than those derived from observations, but they are
similar in magnitude. For the WNL stars this is because the Potsdam
atmosphere models provide smaller line luminosities than suggested by
Table \ref{brinchman_lums}. The WNE line luminosities calculated from
the Potsdam atmosphere models do agree in general with the line
luminosities in Table \ref{brinchman_lums}. The largest mismatch is
for the blue WR bump line luminosity.

Table \ref{lineluminosities} lists the mean line luminosities at
different metallicities. The Potsdam Wolf-Rayet atmosphere models only
exist for solar metallicity. However it is possible to use them at
other metallicities because the helium, carbon, nitrogen and oxygen
composition of Wolf-Rayet stars is almost independent of initial
metallicity and is determined by the core nuclear burning
reactions. The slight changes that do occur are accounted for by the
lower iron opacity in the evolution models, making the surface
temperature and radius of the modelled stars greater and smaller
respectively. This biases the population towards earlier Wolf-Rayet
spectra at lower metallicities. However as Table
\ref{lineluminosities} shows at the lowest metallicities we vastly
overpredict the mean line luminosities by using these atmosphere
models unaltered. Therefore we adapt the scheme of \citet{brinchmann1}
in that below one-fifth solar metallicity we use the reduced line
luminosities for WNL stars and also reduce the line strengths of the
HeII and WR blue bump in the Potsdam atmosphere models by a factor of
a fifth as indicated from the observations of \citet{crowther2}. This
leads to a closer match to the observed mean line luminosities. We
were uncertain whether to use the reduced line strengths at a
metallicity of $Z=0.004$ or not. We calculated a third set of model
spectra with the line luminosities at this metallicity reduced by an
intermediate value of three fifths. In Table \ref{lineluminosities} we
see that the resulting mean line luminosities are only slightly less
than those at the higher metallicity of $Z=0.008$. It is more physical
to expect a gradual reduction in line luminosities with metallicity
rather than a sharp drop. Therefore in the rest of this paper we alter
the line strengths at $Z=0.004$ by three fifths in the Potsdam models
and use the mean of the values listed in Table \ref{brinchman_lums}
for the WNL stars not covered by the Potsdam models.

The second empirical input into the stellar spectra is to take account
of Of stars. For those we again use the method of \citet{brinchmann1},
enhancing the equivalent width of certain emission lines in the O star
spectrum when the gravity of the O star is less than that of Of stars
as described by \citet{1999ApJS..123....3L}. As discussed by
\citet{brinchmann1}, Of stars must be accounted for to create an
accurate spectrum at low metallicity. These are the most luminous type
of O star and have broad emission lines, but differ from Wolf-Rayet
stars. Therefore when the surface temperature of a model is greater
than 33kK and the gravity is less than $3.676 \log(T_{\rm eff}/{\rm
  K}) -13.253$, we supplement the HeII emission lines at 1640\AA\ and
4686\AA\ to produce line luminosities of $20$ and $1.6 \times
10^{35}{\rm ergs \, s^{-1}}$ respectively.

\begin{table}
\caption[]{Input emission line luminosities used for WR stars with $4
  \le \log(T_{\rm eff}/{\rm K}) \le 4.45$ as discussed in Section
  \ref{sec:model_atmospheres}. Line strengths are given in units of
  $10^{35}\, {\rm ergs \, s^{-1}}$.}
\label{brinchman_lums}
\begin{tabular}{lcccc}
\hline
\hline
             & WNL   & WNL  & WNE  & WNE \\
Metallicity  & He(II) & Blue Bump  & He(II) & Blue Bump\\
\hline
\hline
$<0.2Z_{\odot}$   & 43  &  8.3  &  17  & 1.7  \\
$\ge 0.2Z_{\odot}$& 247 &  31   &  84  & 8.4 \\
\hline
\hline
\end{tabular}
\end{table}

\begin{table}
  \caption[]{Mean emission line luminosities predicted by our models
    for WR stars, in units of $10^{35}\, {\rm ergs \,
      s^{-1}}$. Metallicities marked by a single asterisk have had the
    input line luminosities reduced by a fifth and metallicities marked
    by a double asterisk have the input line luminosities reduced by a
    factor of three fifths as discussed in Section
    \ref{sec:model_atmospheres}. The values given in parentheses are
    the mean line luminosities of WR stars where only the Potsdam
    model atmospheres have been used.}
\label{lineluminosities}
\begin{tabular}{lcccc}
\hline
\hline
             & WNL   & WNL  & WNE  & WNE\\
Metallicity  & He(II) & Blue Bump  & He(II) & Blue Bump\\
\hline
\hline
Single\\
0.001*&       32 (8)   &     3 (4)   &    35   &     0.04   \\
0.001 &      198 (103)   &    26 (19)   &   139   &     0.1    \\
0.004*&       42 (20)   &     5 (4)   &    48   &     1    \\
0.004**&      128 (70)   &    16 (10)    &   98   &     2    \\ 
0.004 &      214 (121)   &    28 (16)   &   148   &     2      \\
0.008 &      202 (105)   &    26 (14)   &   122   &     8      \\
0.020 &      158 (100)   &    22 (13)   &   100   &    17      \\
\hline
Binary\\
0.001*&       48 (14)   &     6 (3)   &    28   &     0.1   \\
0.001 &      218 (74)  &    28 (9)   &   102   &     0.4    \\
0.004*&       46 (13)   &     6 (2)   &    26   &     0.6    \\
0.004**&      148 (38)  &   19  (5)   &   58     &     1   \\
0.004 &      206 (63)  &    27 (8)   &    89   &     2      \\
0.008 &      182 (49)   &    24 (6)   &    67   &     4      \\
0.020 &      145 (52)   &    20 (8)   &    58   &     7      \\
\hline
\hline
\end{tabular}
\end{table}

\subsection{Producing a total synthetic population spectrum}
\label{sec:synthesis}

The procedure outlined above yields a synthetic spectrum appropriate
to each timestep of a stellar evolution model.

To construct a synthetic spectrum for the population as a whole we
combine the spectra for models of different initial mass, scaling the
contribution of each by its bolometric luminosity and weighting by the
timestep and initial mass function. Thus to scale the spectra for a
specific population one has simply to select the required total
initial mass and determine the star formation history.  We use two
sets of model: single stars and binary stars. Because we are using
detailed binary models we can follow the effects of binary evolution
at every timestep. This is particularly advantageous during
mass-transfer events when the hydrogen envelope is being removed. We
bin the spectra in time by $\log({\rm age})$ with bins 0.1 dex
wide. The youngest age we consider is 1 Myr.

When a spectrum is added into the total synthetic population we weight
its contribution by a Salpeter IMF with $\alpha=-2.35$. We calculate
the total stellar mass but assuming minimum and maximum initial masses
of 0.1 to $120 \, {\rm M}_{\odot}$. All the model populations
presented below have an initial mass of $10^{5} \, {\rm
  M}_{\odot}$. For the binary populations we use the same IMF to
determine the mass of the primary star and set the parameters of the
secondary as discussed in section \ref{sec:binary-mod}. The total mass
of primaries and secondaries is taken to be the same as for the single
stars, that is $10^{5} \, {\rm M}_{\odot}$.

The result of this process is a totally synthetic simulated spectrum
for a young massive star population over any timescale and metallicity
required, with any star formation history.

We note that shifting from the Salpeter IMF to the Kroupa IMF used by
\citet{EIT} has no measurable effect on the He\,{\sc ii} line. Varying
the IMF by $\pm 0.35$ can alter the C\,{\sc iv} EW by 10 percent but
only at ages below 2Myrs. At older ages the effect is negligible. A
larger effect is found on the WR features in the optical spectrum. We
find that the optical WR features here can vary by up to 5 percent as
we alter the number of low mass main sequence stars contributing to
the optical continuum relative to the number of WR stars. Such small
changes in the IMF also have little effect on the ratios given in
\citet{EIT}.

\subsection{Taking account of nebular emission and other details}
\label{sec:taking-acco-nebul}

One final detail in our spectral synthesis is to include the
contribution from nebular emission. In star-forming galaxies,
interstellar gas is ionised by the stellar continuum emitted blueward
of 912\AA, and upon recombination it emits a nebular
continuum. Neglecting this emission would lead to an incorrect
estimate of the equivalent widths of emission lines and incorrect
broad-band colours \citep{zack,molla}. We use the program
\textsc{Cloudy} to produce a detailed model of the output spectrum
from our stellar spectra. We give the details of the \textsc{Cloudy}
models we use below. The model output is sensitive to the chosen inner
radius and composition of the gas used in the code. The details of our
illustrative nebular emission model are as follows:

\begin{itemize}
\item \texttt{metals ($Z$/0.02) linear}
\item \texttt{element scale factor hydrogen ($(0.75-2.5Z) /0.7$) linear}
\item \texttt{element scale factor helium ($(0.25+1.5Z) /0.28$) linear}
\item \texttt{hden 2 log constant density}
\item \texttt{covering factor 1.0 linear}
\item \texttt{filling factor 1.0 linear}
\item \texttt{sphere}
\item \texttt{radius 1.0 log parsec}
\item \texttt{iterate}
\item \texttt{set temperature floor 1000}
\item \texttt{stop temperature 100K}
\item \texttt{stop efrac -2}
\end{itemize}

The details can be altered to reproduce a more accurate model spectrum
however we only use this simple scheme here to demonstrate how the
synthetic spectrum is affected by nebular emission in Section 3.1. We
do not attempt to model the nebula emission lines when we compare our
models to observed spectra in Section 3.3 below. To match the nebula
emission lines requires altering the input of our \textsc{Cloudy}
model and only tell us the composition of the nebula gas while in this
paper we are primarily interested in the stellar emission. We also
omit dust from our \textsc{Cloudy} models; we note that we are only
concerned with young stellar populations and with the ratio between
line emission and the continuum. Assuming a uniform dust geometry,
these will be suppressed equally leaving the line equivalent widths
unaffected. However as discussed by \citet{charfall} and
\citet{uncertain} for galaxies made up of multiple stellar population
the situation may be more complicated as the birth clouds may disperse
on average after 10 million years and so older populations may be
attenuated less by dust than younger populations and therefore
dominate the observed spectrum.

The theoretical model spectra in their raw state represent a
population in which all the stars are static, lying at precisely the
same redshift or recession velocity, and makes no account for the
velocity dispersion of stars within the observed galaxies. To account
for broadening due to the velocity dispersion of stars within a
typical star-bursting galaxy ($\la 100$\,km\,s$^{-1}$), we convolve
our final spectra with a boxcar function of width 1\AA. For strongly
star-forming galaxies at higher redshifts a higher velocity dispersion
might be appropriate.

\begin{figure*}
\includegraphics[angle=270, width=84mm]{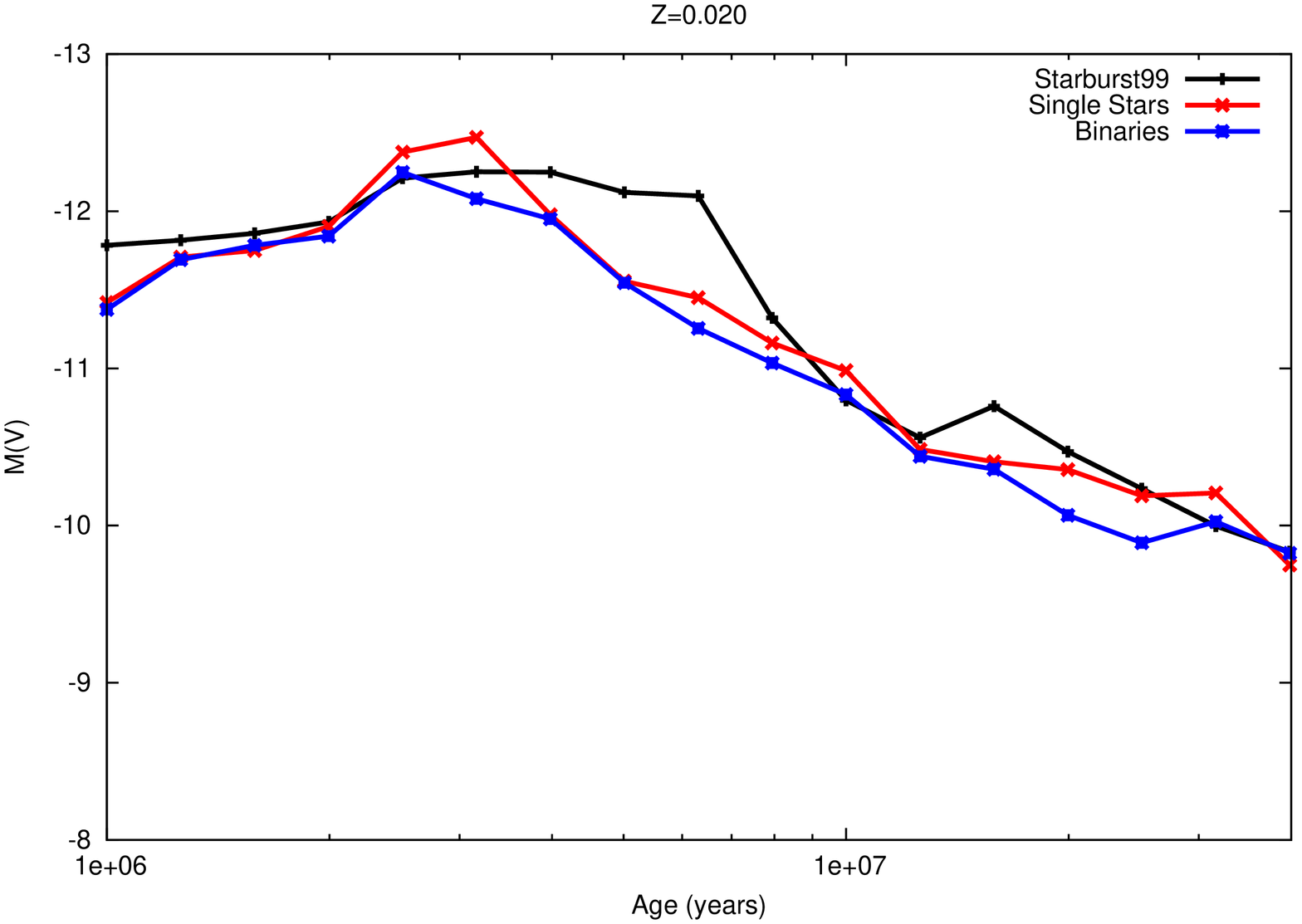}
\includegraphics[angle=270, width=84mm]{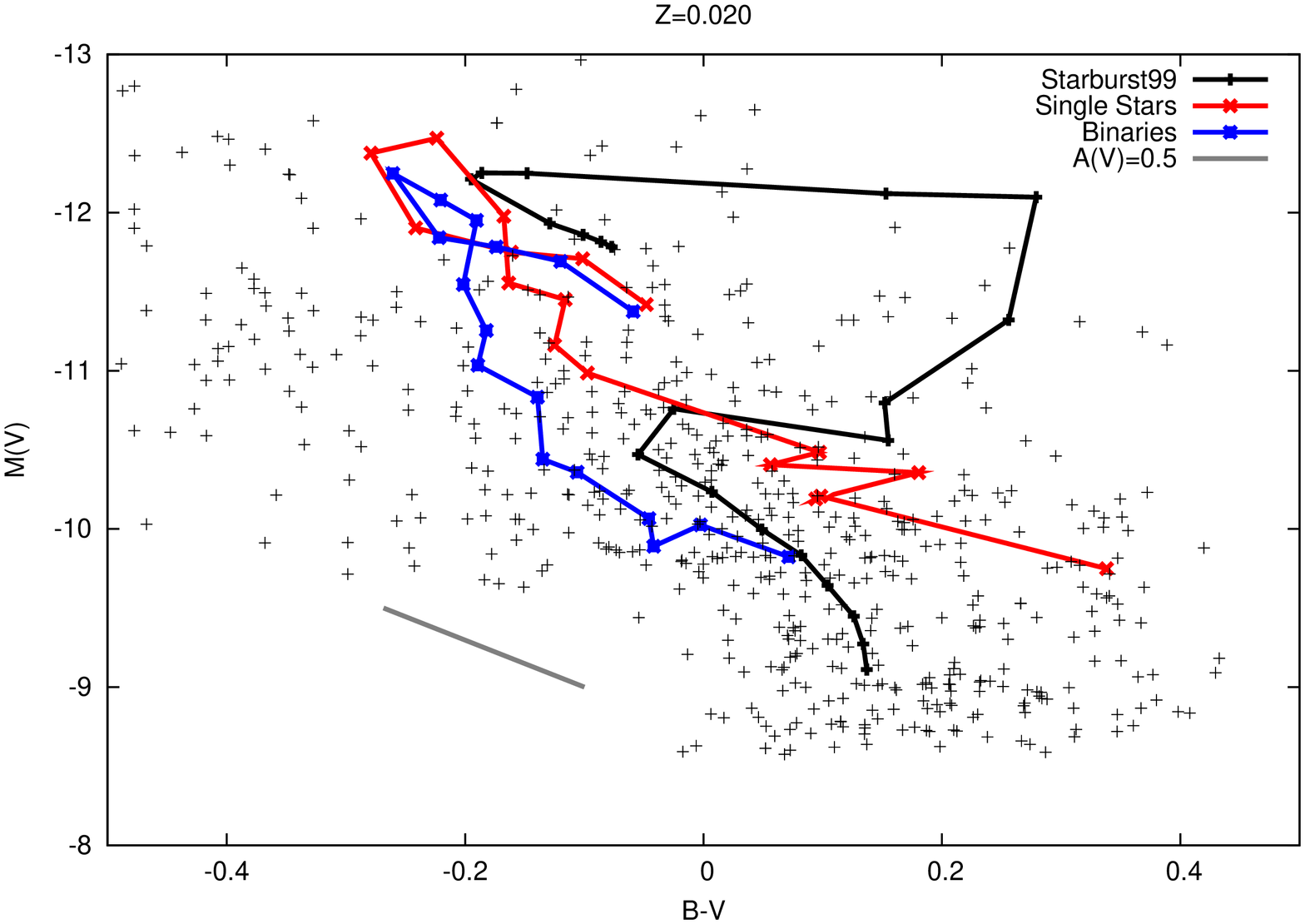}

\includegraphics[angle=270, width=84mm]{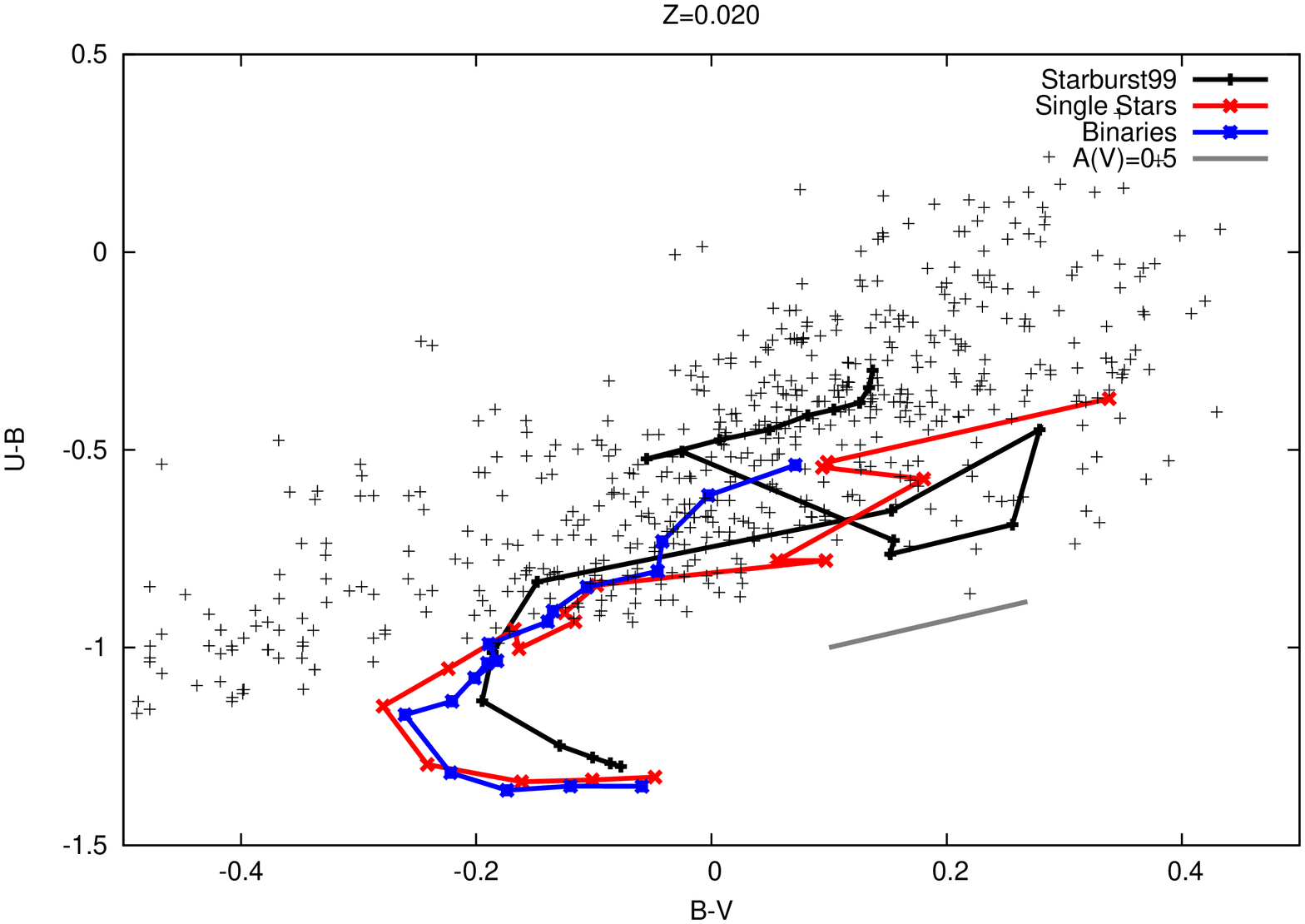}
\includegraphics[angle=270, width=84mm]{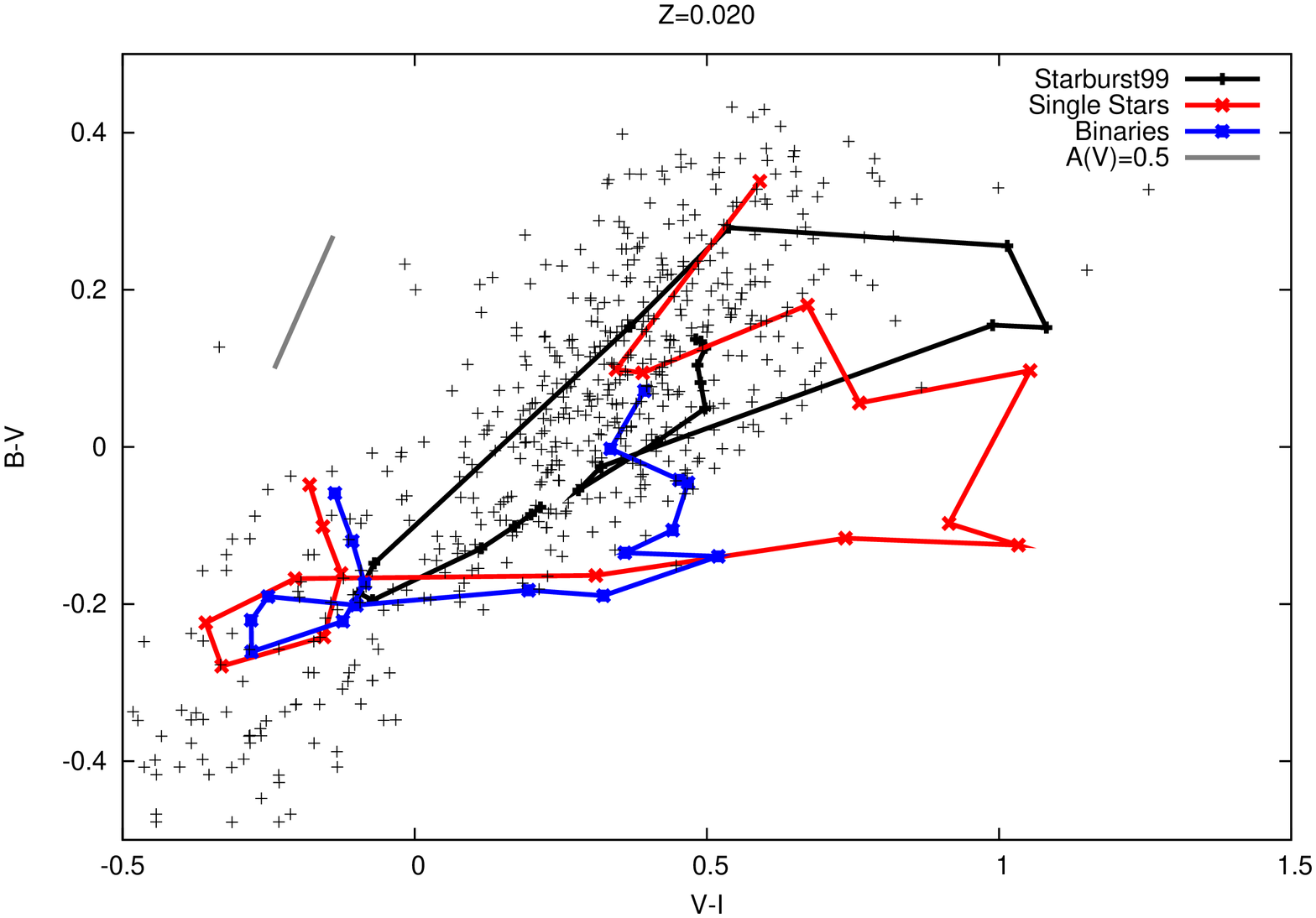}
\caption{The theoretical broad-band colours for a stellar population
  with a total mass of $10^5 M_{\odot}$ compared to observations of
  unresolved young massive clusters in local galaxies from
  \citet{ymclusters}. The theoretical models include a prediction from
  \textit{starburst99}, our single-star population and our binary
  population all including nebular emission at solar metallicity. The
  change in colours produced by extinction of $A_{\rm v}=0.5$ is
  indicated by a vector in each case. Tickmarks on the colour tracks
  indicate age increments of 0.1 dex, from $1$Myr at the bluest B-V
  colours to 40 Myr.}
\label{ymclusters}
\end{figure*}

\section{Verification and local application}
\label{sec:verif-local-appl}

Any new model must be tested against observations. Here we compare our
models to three different sets of observations on sites of
star-formation expected to have WR stars present.

\subsection{Unresolved young massive star clusters  in nearby galaxies}
\label{sec:WR_in_M31}

Before we compare our models to distant galaxies it is sensible to
compare them to nearby unresolved stellar populations. Therefore we
compare our models to a large set of observations of young massive
star clusters compiled by \citet{ymclusters}. They observed a number
of star clusters in nearby spiral galaxies with broad-band
photometry. Since the host galaxies were spiral we assume the
metallicity of the clusters is broadly solar and compare the
observations to colours from our models for a cluster with a mass of
$10^5 \, {\rm M}_{\odot}$. For comparison we have created a similar track from
\textit{starburst99} with the same total stellar mass and initial mass
distribution. This track was calculated using `Starburst99 for
Windows' \citep{starburst4windows} using the same IMF and total mass
as for our models, the standard Geneva solar metallicity stellar
evolution tracks, the Lejeune stellar atmosphere models and the
remaining options at their default values. We plot the results in
Figure \ref{ymclusters}. Our tracks are shorter than the
\textit{starburst99} results as we terminate our simulations at 40
Myrs since we do not currently include stars with initial masses below
$5 \, {\rm M}_{\odot}$ which become important at times later than this.

From Figure \ref{ymclusters} we see that the evolution of colours is
broadly consistent between models and observations. However there are
a number of important differences between the model tracks. These
differences are primarily due to the different stellar models employed
by \textit{starburst99} and the models presented here. Our model
tracks tend to pass through regions of the diagram that contain more
observed clusters although the observations have a large scatter, similar
to the random error of the observations. In B-V the
\textit{starburst99} models tend to have redder colours than our
models at around 10 Myrs. However we see the greatest differences
between the different model tracks in V-I. Our single star population
tracks extend much further into the red than our binary models because
binary models reduce the number of red supergiants and therefore
reduce their contribution in the I band. However, the overproduction
of red supergiants is a feature of all models containing only single
stars (with the main difference being how those supergiants are
characterised in the model population) and emphasises the need for
binary population models.

Our binary models also tend to be bluer in B-V than our single star
models, this is due to the increase in the number of main-sequence
stars at late times due to secondaries accreting mass in binary
interactions. We note that all the models plotted, those presented in
this paper and similar models from \textit{starburst99}, deviate from
the locus of observed clusters in the B-V vs U-B colour plane at early
times ($<5$\,Myr), most likely due to variations in the metallicity of
the clusters away from the Solar composition of our tracks. This
deviation may also be due to relatively simple approximations for the
strong nebular continuum contribution at these times. Models omitting
this nebular contribution, while unphysical, can provide a better
fit to the data in this region, suggesting that more detailed
modelling of the nebular emission may be necessary.

There are a number of factors that we have not included in our model
tracks presented here. For example we have not attempted to fit
absorption from dust in the HII region \textsc{Cloudy} model. This is
because the \textit{starburst99} model track does not include dust,
only nebula continuum emission. In each of the panels we indicate the
reddening direction for an $A_{\rm v}$ of 0.5. We see that in some of
the scatter of the observations could be explained by line of sight
dust.

Given these caveats, we nonetheless find that binary populations have
very different colours to a single star population at certain ages,
and that the observational data at these points is often more
consistent with the binary models with few sources showing V-I
colours, for example, as extreme as those predicted for single star
populations. To achieve this difference between the single and binary
models, binaries with orbital separations between $100$ to
$1000 \, {\rm R}_{\odot}$ must be included. Wider binaries do not interact so
produce results little different from those from single stars. Tighter
binaries tend to experience mergers and so, while evolving as binaries
for some of their lives, they eventually become single stars.

\subsection{Unresolved stellar populations from SDSS}
\label{sec:sdss_gals}

\citet{brinchmann2} recently presented a study of a selection of SDSS
galaxies showing evidence for ongoing massive star formation. They
searched the SDSS DR6 archival spectra to identify those with
Wolf-Rayet features in the optical. The two features used for
identification are known as the blue and red Wolf-Rayet bumps at
approximately 4700\AA\ and 5800\AA.  These star-forming galaxies host
massive stellar populations similar to those incorporated in our
models, and hence provide a good experimental verification of the
predictive power of our models.  In order to construct an appropriate
stellar population in our synthetic spectrum, we assume an
instantaneous burst of star formation and consider its evolution with
time\footnote{Note that this differs from our earlier work
  \citep{EIT}, which considered continuous star formation.}. We
calculate the strength of Wolf-Rayet features both without and with a
nebula emission model created with \textsc{Cloudy} as detailed in
Section \ref{sec:taking-acco-nebul}.  Inclusion of nebular continuum
emission leads to only a slight decrease in the maximum EW allowed by
up to 10 percent.

\begin{figure}
\includegraphics[angle=0, width=84mm]{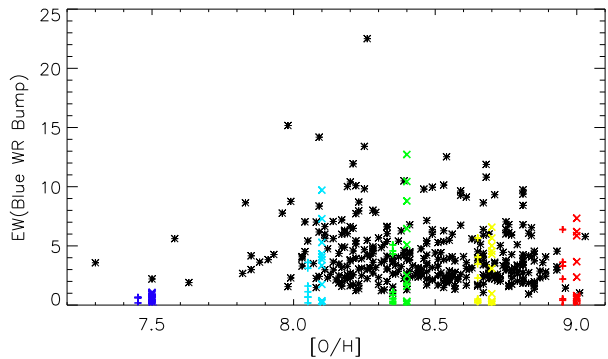}
\caption{The EW of the blue WR bump measured in SDSS galaxies and
  corrected for nebular emission by \citet{brinchmann2} compared to
  predicted EW from our models without nebula emission. Small black
  symbols indicate observations. The models are represented by $+$ for
  single-star populations and $\times$ for binary populations. The
  colour coding for the models are red: twice solar, yellow: solar,
  green: two-fifths solar, cyan: one-fifth solar and blue:
  one-twentieth solar. The vertical spread is due to the change of the
  EW with age of the stellar population.}
\label{final1}
\end{figure}

\begin{figure}
\includegraphics[angle=0, width=84mm]{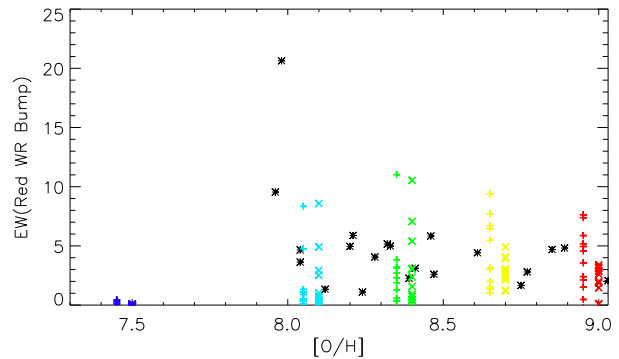}
\caption{Similar to Figure \ref{final1} but for the EW of the red
  Wolf-Rayet bump.}
\label{final2}
\end{figure}

Here we calculate predicted values for the strength (equivalent width;
EW) of Wolf Rayet features as defined by Brinchmann et al, removing
the contribution of nebular lines to allow fair comparison with the
data. As illustrated by Figures \ref{final1} and \ref{final2} in
general our models reproduce the range of observed equivalent widths
of both features with the exception of a few extreme cases.

\begin{figure}
\includegraphics[angle=0, width=84mm]{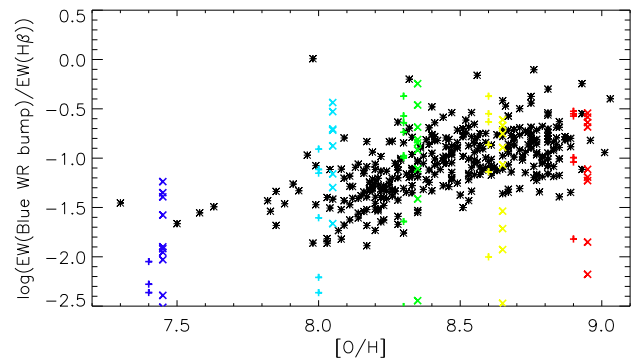}
\caption{Similar to Figure \ref{final1} and \ref{final2} but now for
  the ratio of the Blue WR bump to the H-$\beta$ emission line. The
  H-$\beta$ EW is estimated from our \textsc{Cloudy} and is an upper
  limit as the H-$\beta$ EW can be reduced by decreasing the covering
  factor in the model. Therefore the ratios here are a lower limit as
  decreasing the covering factor will increase the ratio. }
\label{final3}
\end{figure}

For a simple check of our nebula model, Figure \ref{final3} shows a
comparison between the equivalent widths of the Blue WR bump to a
nearby nebula emission line, H-$\beta$. Comparing the ratio of these
EWs will indicate whether we are estimating the strength of the nebula
emission lines correctly relative to the stellar spectrum. We see that
our model predictions of this ratio agree with the range of ratios
from observed galaxies and previous estimates from the binary models
of \citet{vanbev2003}. Binary populations at metallicities below solar
metallicity produce the highest ratio we discuss why this is below. We
note our predicted ratios are a lower limit as the ratio we predict
can be varied by changing the details of our \textsc{Cloudy} model,
especially the covering factor. Decreasing this lets more ionising
photons to escape and therefore decreases the nebula emission features
such as the H-$\beta$ line luminosity without significantly affecting
the blue WR bump EW which is determined by the stellar
population. Thus the vertical spread has degeneracy between age and
covering factor.

In Figure \ref{final7b} we show the strength of the key Wolf-Rayet
emission features in our model spectra as a function of the age of the
stellar population and its metallicity. We see that the spread of
values in Figures \ref{final1}, \ref{final2} and \ref{final3} is due
to a range of ages in the stellar population. Emission features peak
at ages of a few million years for single stars and can be extended to
much later times, up to 10 million years, by the inclusion of
binaries. We note that similar diagrams with a nebula continuum
component would reduce these observed EWs by a small amount. In Figure
\ref{final5} we show how the strength of H-$\beta$ varies with age. We
see that binary models exhibit H-$\beta$ to later ages. This is why at
solar metallicity in Figure \ref{final3} at Solar metallicity single
stars have a greater maximum ratio than binary stars, a smaller
H-$\beta$ EW gives a larger ratio rather than great blue WR bump EW.

In Figure \ref{final7b} there are large peaks in the strength of HeII
and the blue WR bump at around 10 million years. These peaks are due
to WNL stars formed from binary evolution. If these stars were in
binaries they would be red supergiants. Their WNL phase has a long
lifetime due to weak stellar winds at lower metallicities, so it takes
longer for stellar winds and binary evolution to remove their hydrogen
envelopes than at higher metallicity. Therefore their contribution to
the composite spectrum is greater. Also as seen in Figure \ref{final5}
the H-$\beta$ EW is small therefore artificially boosting the ratio of
the two EW in Figure \ref{final3}. If we only consider the ratio at
ages less than $10^{6.8}$ years we do not achieve the highest ratios
in Figure \ref{final3} and the ratio decreases with the same trend
indicated by observations.

The reasons why such low metallicity systems with large ratios are not
observed are related to the work of \citet{charfall} suggesting the
gas in a massive cluster disperses by around 10 million years. With
little gas surrounding the stars most ionising photons will escape
producing little observation H-$\beta$ emission, effectively
decreasing the covering factor. In the sample of SDSS galaxies we use
only a few galaxies have H-$\beta$ EW less than 10\AA. If the covering
factor decreases with ages as suggested by the work of
\citet{charfall}, our predicted H-$\beta$ EW will drop below this
value. Therefore it becomes less likely to observe the large ratios of
blue WR bump EW to H-$\beta$ EW we predict in Figure \ref{final3}.

\subsection{Comparison to the spectra of Tol A \& B}
\label{tolab}

A further comparison between our models and the integrated spectra of
young stellar populations can be made using star forming regions with
good spectral and photometric coverage.  \citet{sidoli} observed the
massive stellar population in the giant HII region Tol89 in
NGC5398. The published spectra of these regions span from the UV
(useful for comparison to high redshift sources) to the optical
(easily observed in more local galaxies). \citet{sidoli} attempt to
fit the observed Wolf-Rayet lines with a \textit{starburst99}
model. They find that for the two knots of star formation, A \& B, the
ages are $4.5\pm1.0$ and $<3$ Myrs respectively with an
LMC-like metallicity, roughly two-fifths solar, preferred.

\begin{figure}
\includegraphics[angle=0, width=84mm]{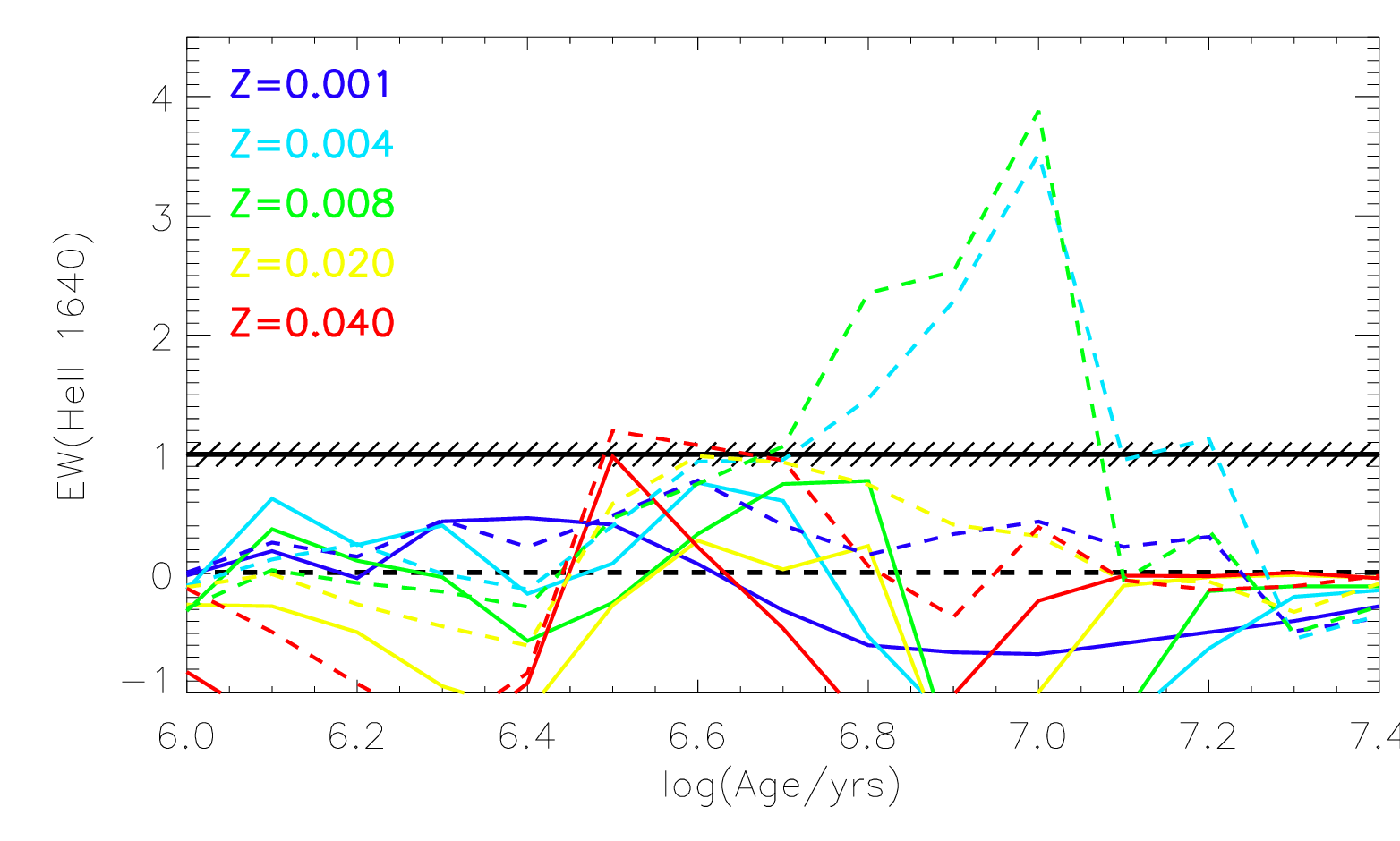}
\includegraphics[angle=0, width=84mm]{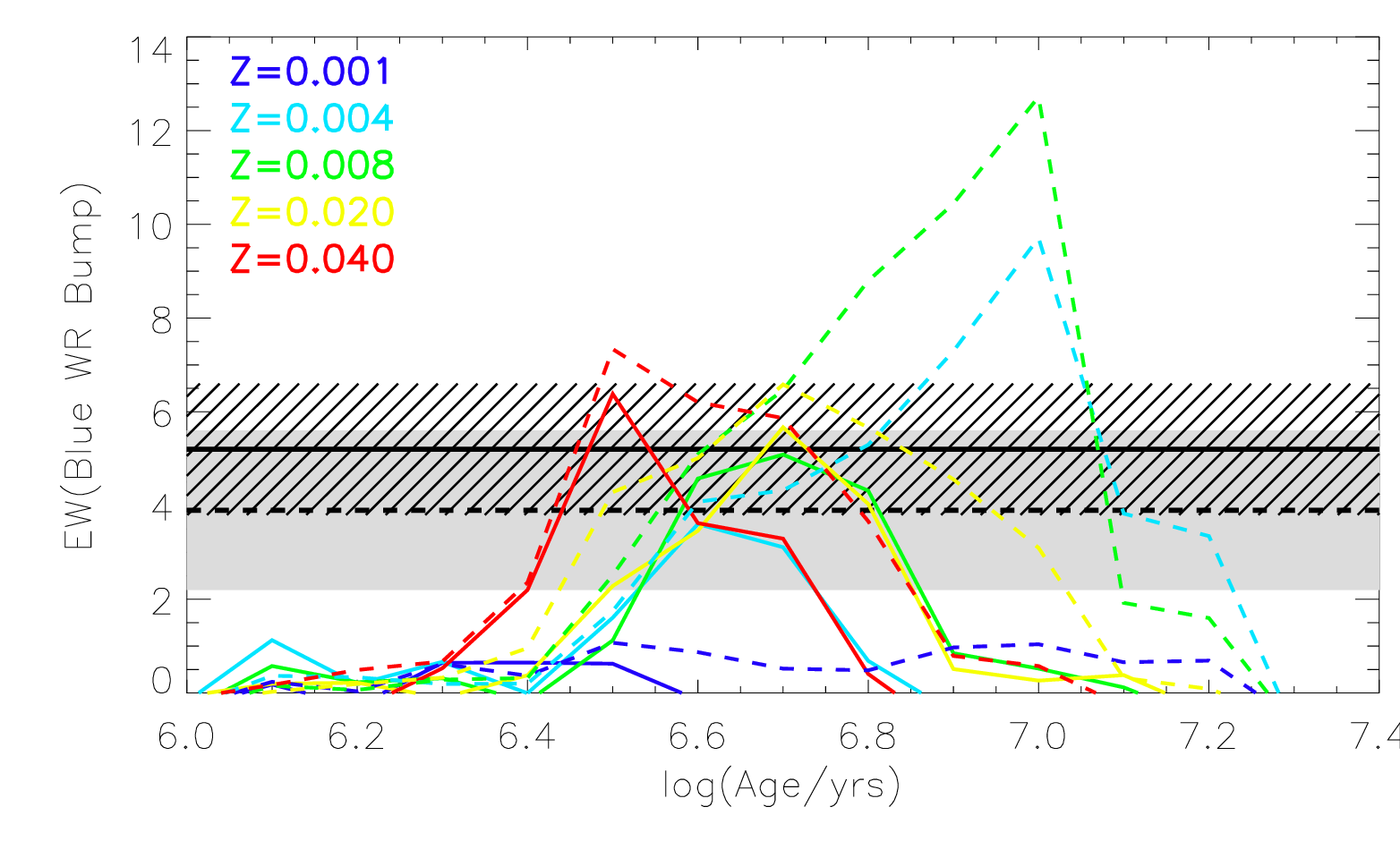}
\includegraphics[angle=0, width=84mm]{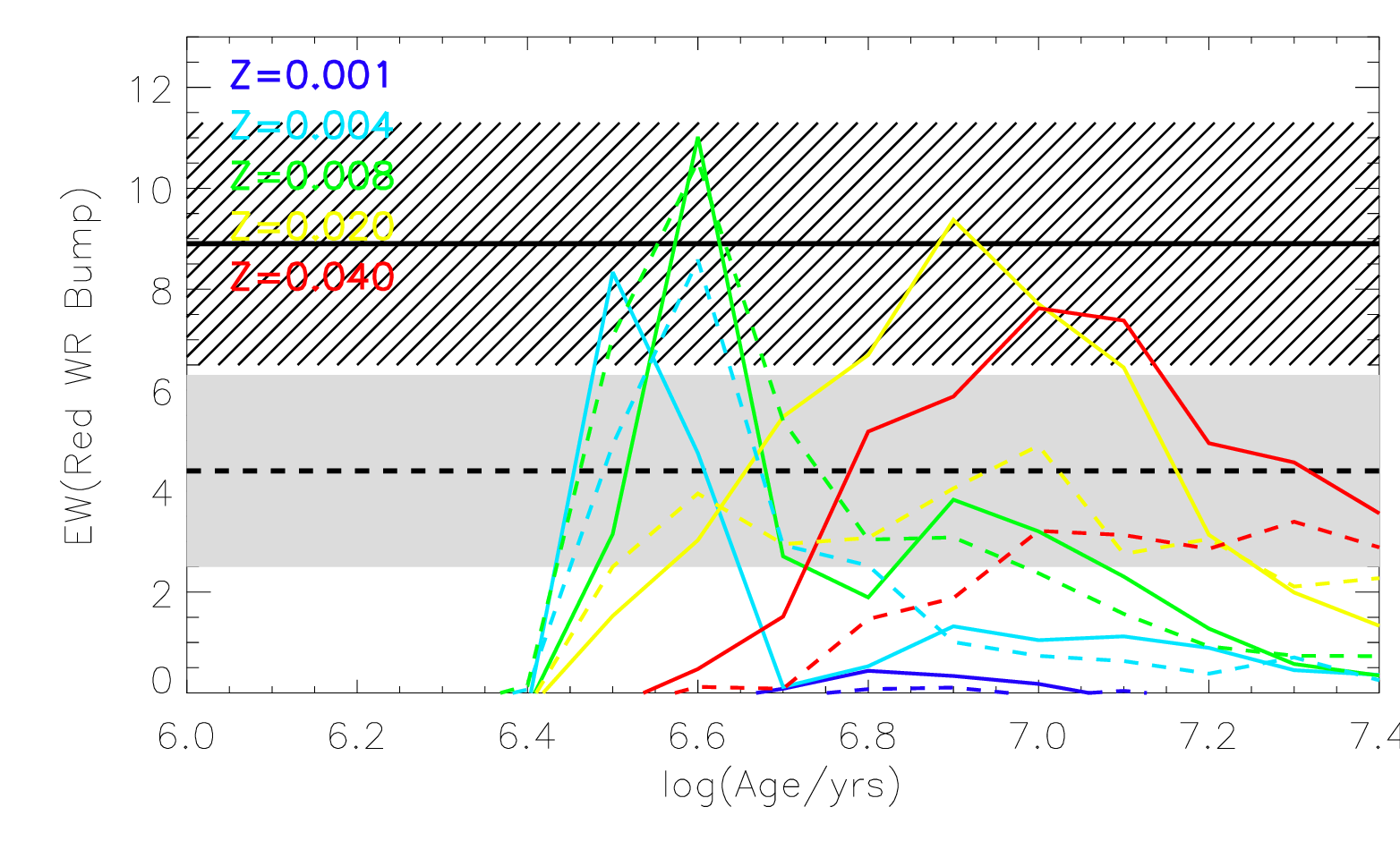}
\caption{The EW of the main Wolf-Rayet spectral features measured from
  our model spectra as a function of age and metallicity. Solid lines
  are for single stars models, the dashed lines are for binary star
  models. The colour scheme for the metallicities is given in Figure
  \ref{final1}. The horizontal black lines represent the measurement
  of these lines for Tol A \& B. The solid line is for Tol A with the
  uncertainty shown by the striped region. The dashed line is for Tol
  B with the grey region showing the uncertainty. These synthetic
  spectra do not include a nebular emission model.}
\label{final7b}
\end{figure}

\begin{figure}
\includegraphics[angle=0, width=84mm]{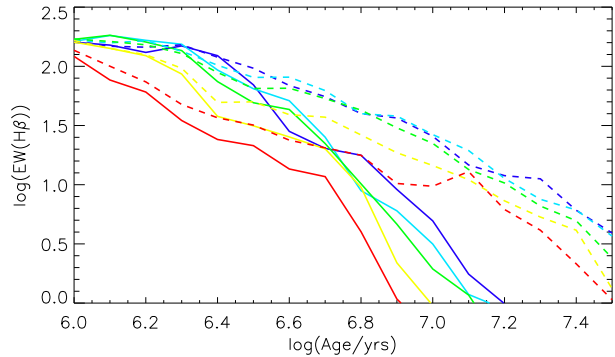}
\caption{Similar to Figure \ref{final7b} but now for the EW of the
  H-$\beta$ emission line. The H-$\beta$ EW is estimated from our
  \textsc{Cloudy} models and is an upper limit as the H-$\beta$ EW can be
  reduced by decreasing the covering factor in the model. While we do
  not use this EW to determine an age for Tol-A \& B we note that the
  EW from the observed spectra are $\log ( \rm{EW(H-} \beta))= 1.6$ and
  2.2 respectively which broadly agree with our predicted H-$\beta$ EW
  at the ages we derive from the stellar emission lines.}
\label{final5}
\end{figure}

We attempt to fit the same spectra with our models. To do this we
first measure the equivalent widths of the ultraviolet HeII
$\lambda$1640\AA\ emission line and the Red and Blue Wolf-Rayet bumps
in an identical manner on both the observed and model spectra. The
HeII EW is calculated by normalising the spectra using the continuum
points identified by \citet{2004ApJ...615...98R} and estimating the EW
over the wavelength range 1635 to 1650\AA. We calculate EWs for the WR
bumps as in Section \ref{sec:sdss_gals} but studying the spectrum to
remove any (relatively narrow) nebula lines. The UV spectrum of Tol A
is a stack of the spectra from 4 individual knots of star
formation. We find one of these knots, A4 as identified by
\citet{sidoli} has a very narrow nebula emission of HeII at
1640\AA. This dominates the EW of the A4 knot and has an equivalent
width of 0.2\AA\ in the Tol A stack. Nebula HeII emission indicates a
very hard ionising radiation field in this cluster. It has been
suggested such lines indicate the presence of population III stars
\citep{2003A&A...397..527S}. However in this case the hard ionising
field is more likely to come from some other source given the
relatively high metallicity and zero redshift. For example that knot
is old enough that there could be an accreting black hole or neutron
star present producing harder radiation. We have not yet attempted to
include such emission in our models yet. Therefore we have calculated
the EW of the HeII line of Tol A ignoring the A4 knot.

We plot our model line strengths against age in Figure \ref{final7b}
compared to the observed EW of features in the Tol A \& B regions. We
find that single star and binary models are both able to reproduce the
observed EW in these features simultaneously at around $10^{6.5}$
years. We also show how show how the H-$\beta$ EW varies with age in
Figure \ref{final5}. While we find that the EW broadly agrees with
those observed for Tol A \& B we do not use them to derive the ages of
the stellar population.

The best fit metallicity is an LMC-like metallicity of $0.4{\rm
  Z}_{\odot}$. The age of Tol A is constrained to be from $10^{6.5}$
and $10^{6.7}$ years. At ages outside this range the stellar features
are too weak to agree with the observed values, or for the binary
models, too strong. For Tol B the age is constrained to be less than
$10^{6.4}$ years.

\begin{figure*}
\includegraphics[angle=0, width=84mm]{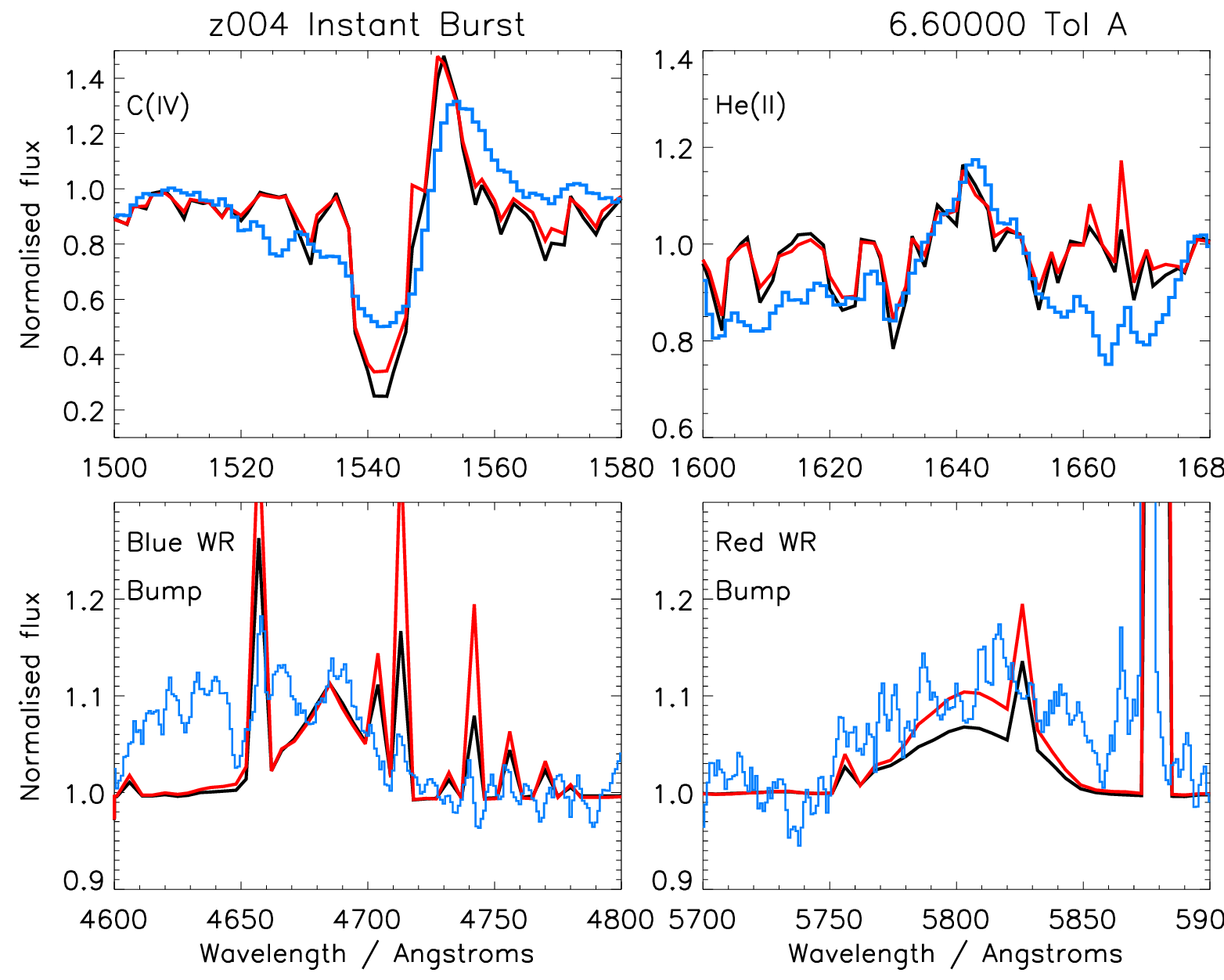}
\includegraphics[angle=0, width=84mm]{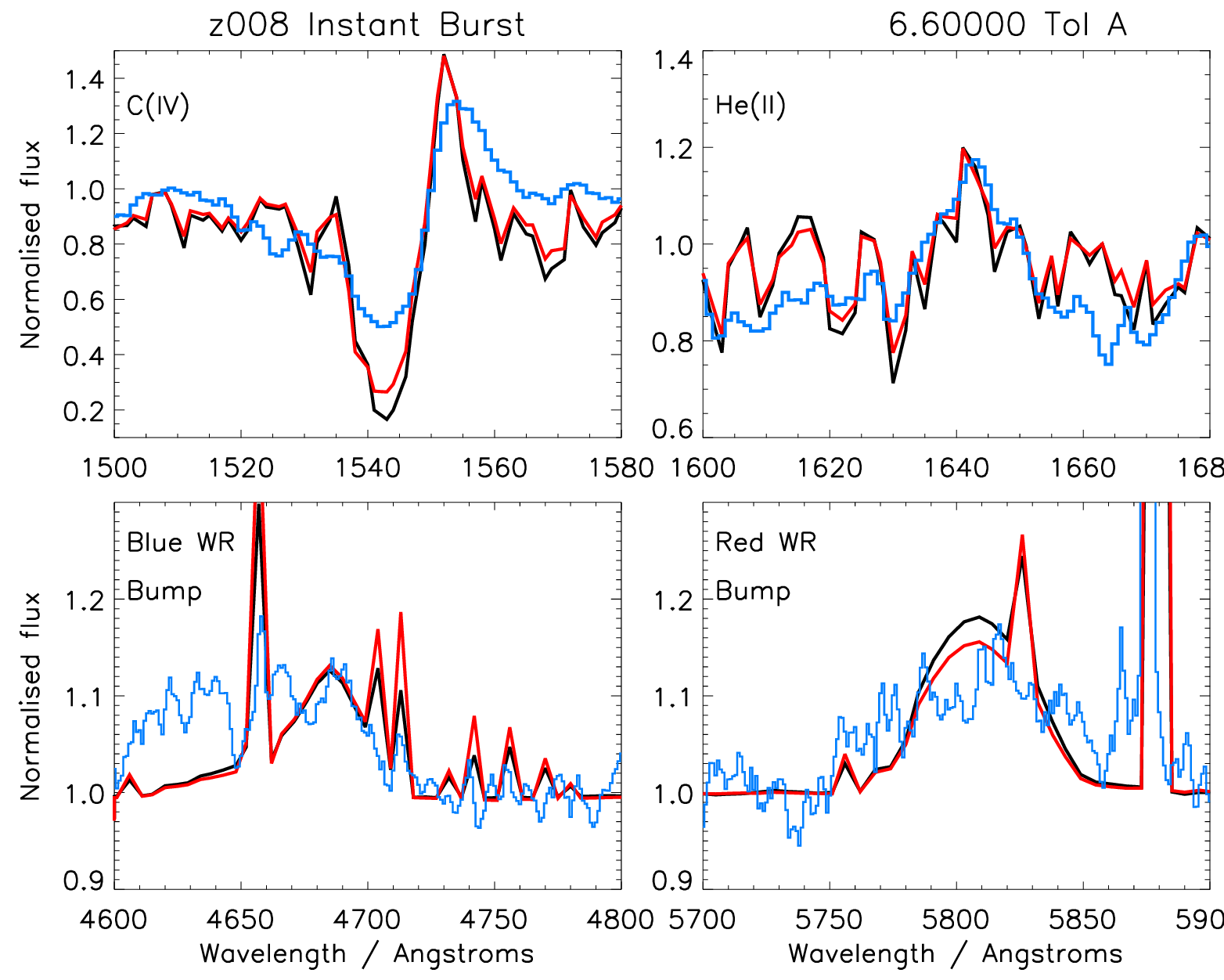}
\caption{Our best fitting spectra for Tol A both for an age of
  $10^{6.6}$~years. The left panel is for a SMC-like metallicity,
  $Z=0.004$. The right panel is for an LMC-like metallicity,
  $Z=0.008$. The observations are shown in blue. The single-star
  models are shown with solid black lines while the binary models are
  the solid red lines.}
\label{final8a}
\end{figure*}

\begin{figure*}
\includegraphics[angle=0, width=84mm]{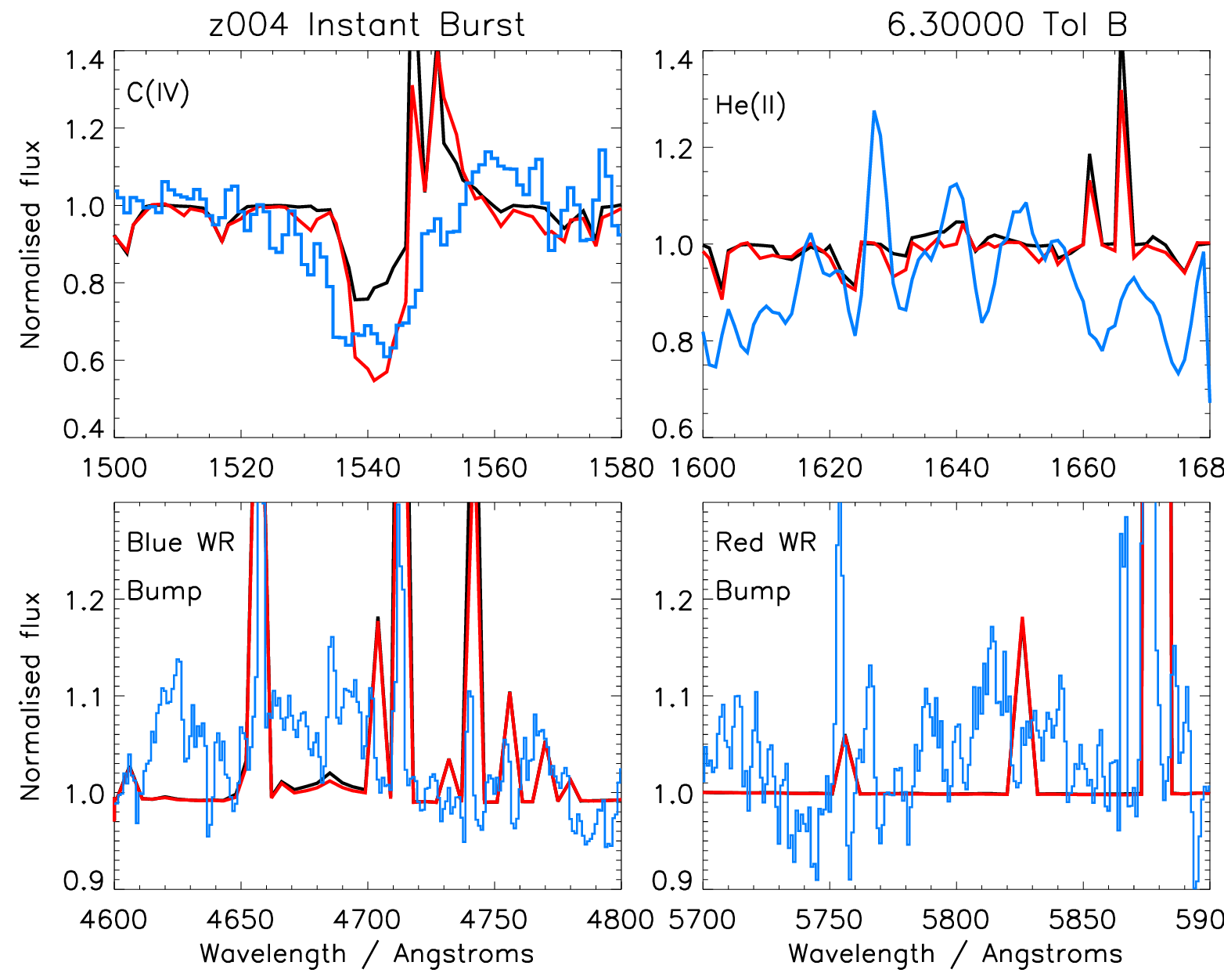}
\includegraphics[angle=0, width=84mm]{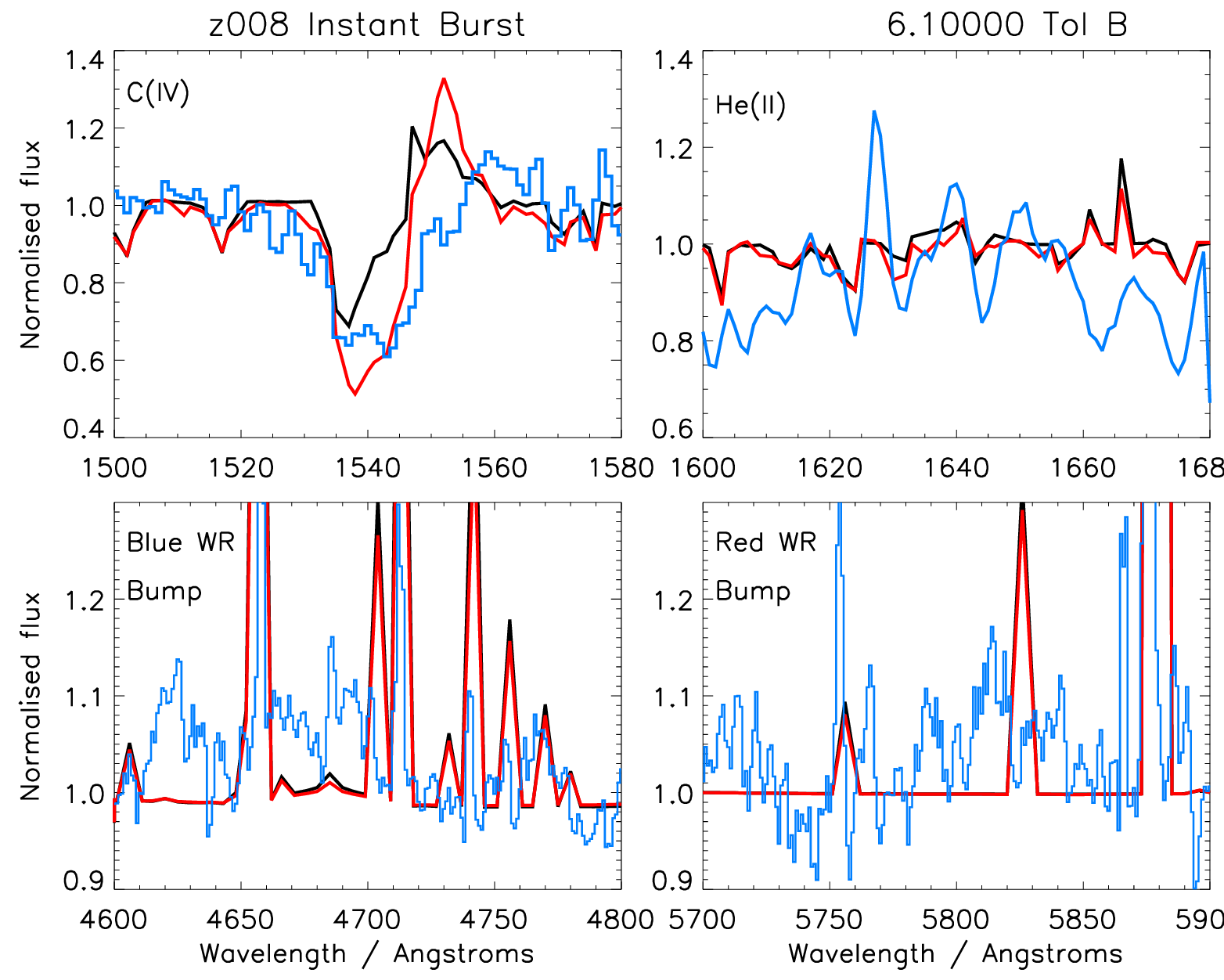}
\caption{Similar to Figure \ref{final8a}, our model spectra for Tol B,
  the left panel is for a model with SMC-like metallicity and an age
  of $10^{6.3}$~years and the right panel is for a model with LMC-like
  metallicity at an age of $10^{6.1}$~years.}
\label{final8b}
\end{figure*}

To verify the best fit ages and metallicities we show a more detailed
comparison between these models and the observed spectra for Tol A and B in
Figures \ref{final8a} and \ref{final8b} respectively, considering the
three key WR diagnostic features and also the ultraviolet CIV
absorption/emission feature at 1550\AA, which is sensitive to the
presence of massive stars and in particular winds driven from O-stars.
For Tol A, we find reasonable visual fits between $10^{6.5}$ and
$10^{6.6}$ years at both SMC ($0.2{\rm Z}_{\odot}$) and LMC
metallicities. The CIV line requires younger ages for the system than
the WR features provide. For Tol B we find any model with negligible
HeII emission provides a reasonable fit to the data. However models
with ages of $10^{6.4}$ years and older tend to overpredict the CIV
emission line relative the observed spectrum. Also for the blue and
red bumps there are no distinctive broad emission lines and the
spectrum is dominated by narrow emission lines. This again suggests a
young population.

Our derived ages, both from matching the spectra by eye and from the
line equivalent widths, are 4 Myrs for Tol A and $<2.5$Myrs for Tol
B. These agree with the ages of \citet{sidoli}. However our ages are
on the young side of the ages they derived.

By eye the model agreement is good, especially in light of the general
models employed, which are not fine-tuned to any great degree. The
line profiles and strengths of ultraviolet spectral features are
generally less well fit than the optical features. There are several
ways in which our models might be fine tuned to fit the observational
data in this case. Firstly, it may be possible to refine our model for
nebular continuum contribution to the integrated spectrum. At present,
we consider a basic model as discussed above and parameters could be
adjusted to gain a better fit for this star-forming region. Secondly,
it may be possible to include the effects of material external to the
system in addition to gas and dust local to the star-forming region
which is modelled using \textsc{Cloudy}. Thirdly, we only use a sparse
grid of O-star models rather than a dense grid of atmospheres and it
may be possible to improve upon this in future
\citep{2004ApJ...615...98R}. Fourthly, we have assumed an
instantaneous burst. A better fit might be achieved by allowing
multiple star formation episodes with slightly different ages. Fifthly
we have not considered the effects of stellar rotation on the
evolution or the spectra. This will have two effects on our
results. It would extend the main-sequence life-time and would also
rotationally broaden the CIV line further, particularly if the stars
rotate at velocities above $200 {\rm km \, s^{-1}}$. This would
potentially improve the fit between models and observed data. We
speculate that study of the CIV profile could potentially give a way
to evaluate the importance of rotation versus binarity in stellar
populations. Finally, some WNL stars have CIV in emission. We have not
included any such CIV emission in the WNL stars that are not
represented by the Potsdam model atmospheres. Including an empirical
correction for such emission in these models may broaden the emission
component of the CIV line and simultaneously decrease the absorption
component. However we do not choose to introduce arbitrary line
emission here.


\section{Discussion and Conclusions}
\label{sec:discussion}

In this paper we set out to describe the construction of model stellar
populations incorporating massive stars and massive stellar binaries,
and then the synthesis of spectra for this population. We have
compared our model spectra to observations of comparable unresolved
stellar populations and found agreement is fair over a large range of
wavelengths, from the UV to the near infra-red.

In general our population synthesis including binary models predict
that such systems have less emission at long wavelengths around the I
band. This is because binary interactions remove the hydrogen envelope
of some red supergiants to form Wolf-Rayet stars. These WR stars then
lead to more blue colours in B-V and V-I broad-band colours and a
larger UV flux. The latter increases the timespan over which nebula
emission is important to the evolution of stellar populations from
6 Myrs to 20 Myrs. Another binary effect is that it is more likely that
strong WR emission lines are observed since binary interactions tend
to spread WR emission features over a longer timespan. For single
stars the WR stars all exist over a short timespan so WR features are
only present for a short period of time. This suggests that ages
derived from Wolf-Rayet features to date may have been systematically
underestimated.

It may be somewhat surprising that by including binary evolution we
find no great difference from predictions from \textit{starburst99}, a
code based on single-star evolution alone. However the stellar
evolution models used in the \textit{starburst99} model present in
Section \ref{sec:WR_in_M31} are nearly two decades old
\citep{oldgeneva}. The important difference between these models and
those used here is that the mass-loss rates have substantially
decreased for OB-stars and WR stars. Therefore with incorrect
single-star mass-loss rates the older models reproduce the
observed stellar population. Our model binary population therefore
should broadly agree with this older code, but now we model the same
magnitude of mass-loss as a combination of lower single-star mass-loss
rates enhanced for some stars by binary interactions which is a
physically reasonable model.

We demonstrate that our code produces a good fit to the observational
data of local stellar populations in which massive stars are
important. However, further refinements are possible and additional
verification data on local star-forming regions would be
welcome. Nonetheless, the stellar models and synthesis code presented
here may now be used as a tool to study stellar populations in a range
of different observational domains and to derive their physical
parameters, as demonstrated by their application to extragalactic
star-forming regions Tol A and B.

\section*{Acknowledgements}

The authors would like to thank the referee Jarle Brinchmann for his
very helpful and constructive comments. The authors also thank Malcolm
Bremer, Max Pettini, Paul Crowther, Stephen Smartt, Nate Bastian and
Norbert Langer for useful discussions.  ERS acknowledges postdoctoral
research support from the UK Science and Technology Facilities Council
(STFC). JJE began this work when he was supported by the award
``Understanding the lives of massive stars from birth to supernovae''
made under the European Heads of Research Councils and European
Science Foundation EURYI Awards scheme which is supported by funds
from the Participating Organisations of EURYI and the EC Sixth
Framework Programme. JJE also acknowledges support from the UK Science
and Technology Facilities Council (STFC) under the rolling theory
grant for the Institute of Astronomy.

\label{lastpage}
\bsp

\end{document}